\documentclass[runningheads]{cl2emult}

\usepackage{makeidx}  
\usepackage{graphicx} 
\usepackage{subeqnar} 
\usepackage{multicol} 
\usepackage{cropmark} 
\usepackage{eso}      
\makeindex            

\newcommand{\lya}{Ly$\alpha$}
\newcommand{\Ha}{H$\alpha$}
\newcommand{\Hb}{H$\beta$}

\def\ltsima{$\; \buildrel < \over \sim \;$}
\def\simlt{\lower.5ex\hbox{\ltsima}}
\def\gtsima{$\; \buildrel > \over \sim \;$}
\def\simgt{\lower.5ex\hbox{\gtsima}}

\begin{document}
\title*{Deep Observations of Lyman Break Galaxies}
%
%
%
%
\titlerunning{Lyman Break Galaxies}
%
\author{Max Pettini\inst{1}
\and Charles C. Steidel\inst{2}
\and Alice E. Shapley\inst{2}     
\and Kurt L. Adelberger\inst{2}
\and Alan F.M. Moorwood\inst{3}
\and Jean-Gabriel Cuby\inst{3}
\and Mark Dickinson\inst{4}
\and Mauro Giavalisco\inst{4}}
%
\authorrunning{Max Pettini}
%
%
\institute{Institute of Astronomy, Madingley Road, Cambridge CB3 0HA, England
\and Palomar Observatory, California Institute of Technology,
     MS 105--24, Pasadena, CA 91125, USA
\and European Southern Observatory, Karl-Schwarzschild-Str. 2, 
     D-85748 Garching, Germany; and Alonso de Cordova 3107,
     Santiago, Chile
\and Space Telescope Science Institute,
     3700 San Martin Drive, Baltimore,
     MD 21218, USA}

\maketitle              

\begin{abstract}
We summarise the main results of recent work on the Lyman
break galaxy population which takes advantage of
newly commissioned instrumentation
on the VLT and Keck telescopes to push the detection of
these objects to new wavelengths and more sensitive limits.
\footnote{\tt Proceedings of the ESO
{\it Deep Fields\/}¥ Symposium. 
To be published in
{\it ESO Astrophysics Symposia}, ed. S. Cristiani
(Berlin: Springer)}
\end{abstract}

\section{Introduction}

Given the specialist nature of this meeting,
I shall concentrate on the most recent results
in our study of Lyman break galaxies (LBGs), rather than
give a broad perspective of the high redshift galaxy
population; up to date reviews of this subject can be found 
in the articles by Dickinson (2000), Ferguson et al. (2000),
and Pettini (2000). As we shall see, the work I am about to 
describe uses `deep' observations of LBGs 
which push the capabilities of present 
instrumentation to its limits.

Since the commissioning of the high resolution,
near infrared spectrographs on the VLT (ISAAC) 
and Keck (NIRSPEC) telescopes in mid-1999, we have been 
engaged in an extensive programme to record 
the rest-frame optical spectra of the brightest
Lyman break galaxies with the aim of complementing
and extending the information provided by their rest-frame
UV spectra on which most of our knowledge 
of these objects rests at the moment.
Apart from the surprises often
associated with opening a new wavelength window, there are
several obvious scientific motivations. 

The luminosity of the
Balmer lines, primarily \Ha\ and \Hb, gives a measure of the star
formation rate which is directly comparable to the values deduced
in local surveys. Furthermore, since the optical emission lines
and the far-UV continuum do not respond to dust extinction to the
same degree, the relative luminosity of a galaxy in these two
tracers of star formation could in principle be used as a
reddening indicator. When integrated over an entire galaxy, the
widths of the nebular lines should reflect the velocity
dispersion of the H~II regions within the overall gravitational
potential so that a kinematical mass may be deduced. This is not
possible in the UV, because the interstellar absorption lines are
sensitive to small amounts of gas accelerated to high velocities
by energetic events such as supernova explosions and bulk
outflows, while the stellar absorption lines from OB stars are
intrinsically broad. Finally, there are well established chemical
abundance diagnostics based on the relative strengths of nebular
emission lines, primarily [O II], [O III] and \Hb. Abundance
measurements are much more difficult in the UV where the more
easily observed interstellar lines are generally saturated so
that their equivalent widths depend mostly on  the velocity
dispersion of the gas and only to a lesser extent on the column
density of the absorbing ions.

With the large
sample of Lyman break galaxies now available (nearly 1000 
spectroscopically confirmed objects),
it is possible to isolate redshifts which
place the transitions of interest in gaps between the strong OH 
lines which dominate the near-IR sky; 
here the background is sufficiently dark for faint
extragalactic work to become possible. 
Figure 1 shows an example of the quality of spectra which can be secured 
with a 2-3 hour integration with NIRSPEC or ISAAC.
\begin{figure}
\vspace*{-3.50cm}
\hspace*{-3.75cm}
\centering
\includegraphics[width=1.2\textwidth,angle=270]{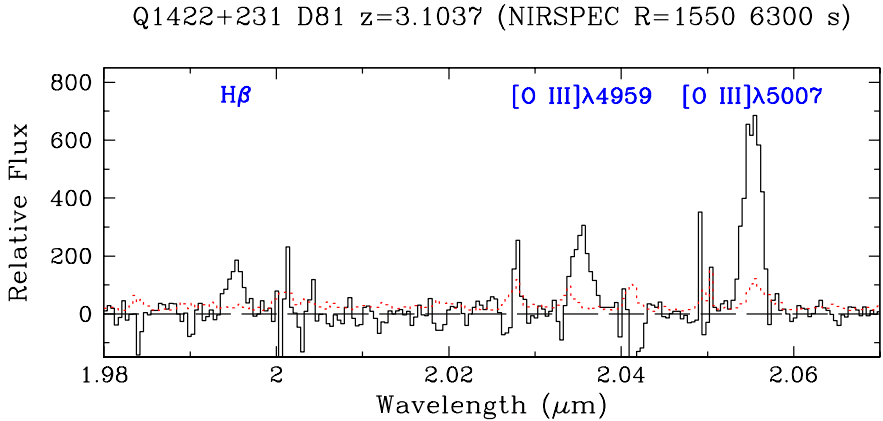}
\vspace*{-4cm}
\caption[]{
Example of a NIRSPEC $K$-band spectrum of a Lyman break galaxy.
The objects targeted in our survey typically have $K = 21$ (on the Vega 
scale) and remain undetected in the continuum, although the 
emission lines are easily seen with 2-3 hour exposures.
The dotted line shows the $1 \sigma$ error spectrum.}
\label{}
\end{figure}
Together with observations from the literature,
our survey consists of 19 LBGs
drawn from the bright
end of the luminosity function, from 
$\sim L^{\ast}$ to $\sim 4\,L^{\ast}$.
Overall we find that galaxies observed are very uniform in their
near-IR properties. The spectra are dominated by the emission lines,
and the continuum is detected in only two objects, one 
of which---West~MMD11---has an unusually red optical-to-infrared
color with (${\cal R} - K_{\rm AB}$)\,$= 2.72$\,.
In all cases [O~III] is stronger than \Hb\ and [O~II].
The line widths span a relatively small range, with 
values of the one dimensional velocity dispersion 
$\sigma$ between 50 and 115\,km~s$^{-1}$.

\section{Star Formation Rates and Dust Extinction}
The star formation rates deduced from the luminosity
of the \Hb\ emission line agree within a factor of $\sim 2$
with the values implied by the continuum luminosity
at 1500\,\AA\ {\it before any corrections for dust extinction
are applied\/} (see Figure 2). 
\begin{figure}
\vspace*{-2.75cm}
\hspace*{-1.75cm}
\centering
\includegraphics[width=1.0\textwidth,angle=270]{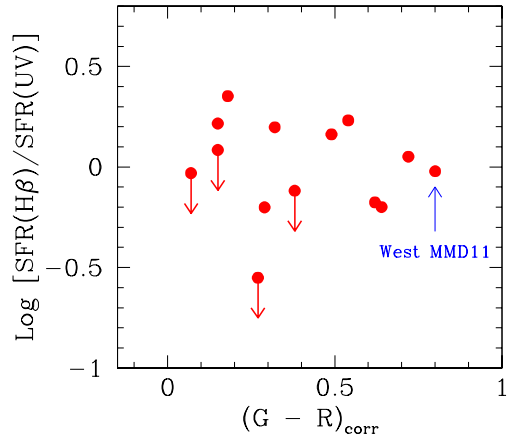}
\vspace*{-3cm}
\caption[]{Comparison between the values of star formation rate
deduced from the luminosities in the \Hb\ emission line
and in the UV continuum at 1500~\AA.  
The color ($G - {\cal R}$) measures the intrinsic 
UV spectral slope after statistical correction
for the \lya\ forest opacity. Note that in the SCUBA source
West~MMD11, which is also the reddest object in the present sample,
the strength of \Hb\ relative to the UV continuum
is typical of the rest of the sample and 
SFR$_{\rm H\beta} \simeq$\,SFR$_{\rm UV}$.
}
\label{}
\end{figure}
There is no trend in the present sample
for the former to be larger than the latter, as may have been expected
from the shape of all reddening curves which rise from the optical
to the UV. Evidently, any such differential extinction
must be small compared with the uncertainties in calibrating
these two different measures of the SFR (Kennicutt 1998; 
Charlot \& Longhetti 2001).
This conclusion is in agreement with the results of similar 
recent studies of UV-selected star-forming galaxies at $z \simlt 1$
(e.g. Flores et al. 1999; Sullivan et al. 2000; Bell \& Kennicutt 2001)
and contradicts the commonly held view that the Balmer
lines are more reliable star formation indicators than the UV 
continuum---from our sample one would obtain essentially the same
star formation rate density using either method.

\section{Oxygen Abundance}
In five cases (four new ones and one previously published)
we attempted to deduce values of the abundance of oxygen
by applying the familiar $R_{23}$ 
([O~II]\,+\,[O~III]/\Hb) method which has 
been extensively used in local H~II regions.
We found that generally there remains a significant 
uncertainty, by up to 1\,dex, in the value of (O/H)
because of the double-valued nature of the 
$R_{23}$ calibrator. Thus, in the galaxies observed
oxygen could be as abundant as in the interstellar medium 
near the Sun, or as low as $\sim 1/10$ solar.
While this degeneracy can in principle 
be resolved by measuring the [N~II]/\Ha\ ratio
(and in the one case where this has proved possible---Teplitz 
et al. 2000---values of (O/H) near the upper end of the range are indicated), 
this option is not normally available for galaxies at $z \simeq 3$
because the relevant lines are redshifted beyond the $K$-band. 
\begin{figure}
\vspace*{-2.75cm}
\hspace*{-2.1cm}
\centering
\includegraphics[width=1.0\textwidth,angle=270]{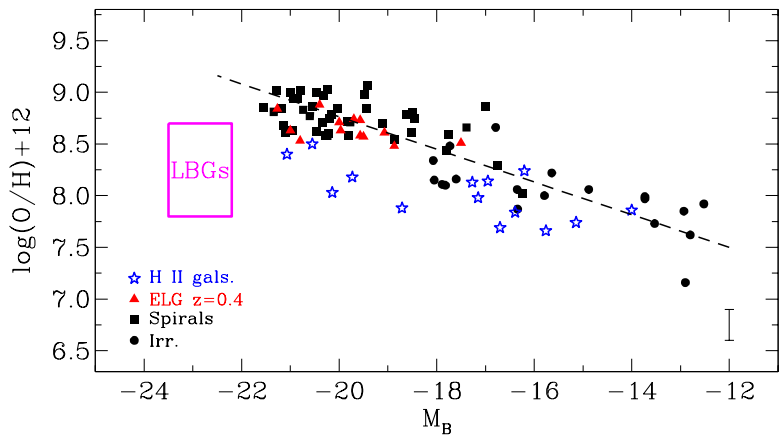}
\vspace*{-3cm}
\caption[]{Metallicity-luminosity relation for local galaxies,
from the compilation by Kobulnicky \& Koo (2001) adjusted to the
$H_0 = 70\,$km~s$^{-1}$~Mpc$^{-1}$, $\Omega_{\rm M} = 0.3$, 
$\Omega_{\Lambda} = 0.7$ cosmology adopted in this work.
The vertical bar in the bottom right-hand corner
gives an indication of the typical error in log(O/H).
In the Sun 12\,+\,log(O/H) = 8.83 (Grevesse \& Sauval 1998).
The box shows the approximate location of the Lyman break galaxies 
in our sample at a median $z = 3.1$\,. 
Like many local H~II galaxies, 
LBGs are overluminous for their metallicity.
The height of the box results largely from the double-valued nature
of the calibration of (O/H) in terms of the $R_{23}$ index; 
the one case where the ambiguity can be resolved
(MS~1512-cB58) lies in the upper half of the box.}
\label{}
\end{figure}
Even so, it is still possible to draw some interesting conclusions.
First, LBGs are definitely more metal-rich than
damped \lya\ systems at the same epoch, which typically have 
metallicities $Z \approx 1/30\,Z_{\odot}$. 
This conclusion is consistent with the view that DLAs are drawn 
preferentially from the faint end of the galaxy 
luminosity function and are not the most actively star 
forming galaxies, as indicated by essentially all 
attempts up to now to detect them via direct imaging.
Second, LBGs do not conform to today's
metallicity-luminosity relation and are overluminous
for their oxygen abundance (see Figure 3).
This is probably an indication that they have
relatively low mass-to-light ratios, as also
suggested by their kinematical masses; 
an additional possibility is that
the whole (O/H) vs. $M_{\rm B}$ correlation shifts
to lower metallicities at high $z$, when galaxies were
younger.

\section{Kinematical Masses}
If the emission line widths reflect the relative
motions of H~II regions within the gravitational potential
of the galaxies, the implied masses are of the order
of $10^{10}\,M_{\odot}$ within half-light
radii of $\sim 2.5$\,kpc. This is likely
to be a lower limit to the total masses of the galaxies
as would be obtained, for example, if we could trace their 
rotation curves. A more serious uncertainty, however,
is the real origin of the velocity dispersions we measure.
\begin{figure}
\vspace*{-2cm}
\hspace*{-0.55cm}
\centering
\includegraphics[width=0.8\textwidth,angle=270]{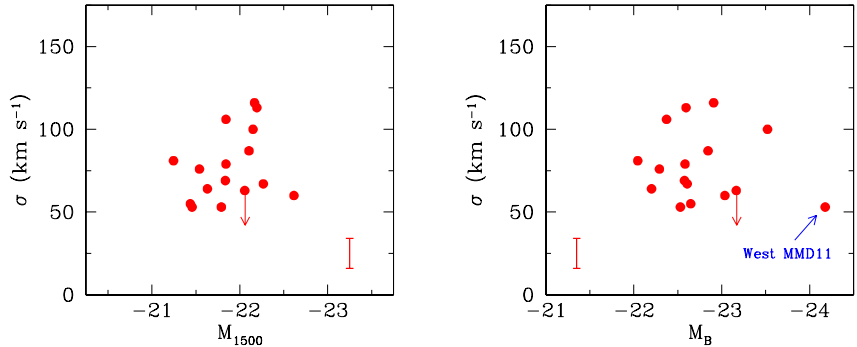}
\vspace*{-2cm}
\caption[]{
One dimensional velocity dispersion of nebular 
emission lines in Lyman 
break galaxies as a function of absolute magnitude in the rest-frame
far-UV (left) and $B$-band (right). The vertical bar shows the typical 
error on the measurements of $\sigma$. Curiously,
the SCUBA source West~MMD11, which has the reddest
(${\cal R} - K$) color in the present sample, exhibits
one of the smallest velocity dispersions.}
\label{}
\end{figure}
We do not see any correlation between $\sigma$ and galaxy luminosity
in either our limited sample (see Figure 4)
nor in an on-going
study by some of us of a much larger 
sample of galaxies at $z \simeq 1$ which span five magnitudes in 
luminosity and yet show very similar line widths to those found here.
In two cases we have found hints of ordered motions
in spatially resolved profiles of the [O~III] lines,
but attempts to use high resolution images to clarify whether they are 
indicative of rotating disks proved to be inconclusive.

\section{Galactic Superwinds}
In all the galaxies observed we find evidence for
bulk motions of several hundred km~s$^{-1}$ 
from the velocities of the
interstellar absorption lines---which are systematically
blueshifted---and \lya\ emission---which is always 
redshifted---relative to the nebular emission lines (see Figure 5).
\begin{figure}
\vspace*{-2cm}
\hspace*{-2.1cm}
\centering
\includegraphics[width=1.0\textwidth,angle=270]{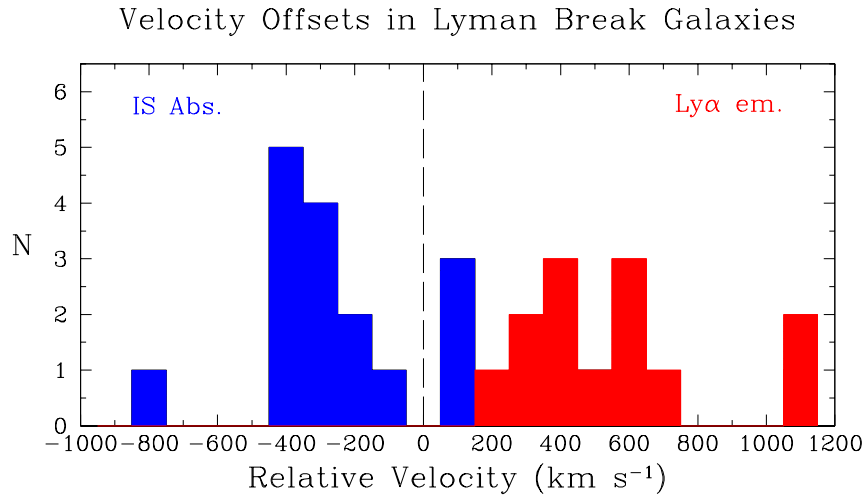}
\vspace*{-3cm}
\caption[]{
Velocity offsets of the interstellar
absorption lines (blue or dark grey)
and of the \lya\ emission line (red or light grey)
relative to [O~III] and \Hb.
Large scale motions of the order of several hundred
km~s$^{-1}$ are indicated by the systematic
tendency for the former to be blueshifted and the latter redshifted
relative to the nebular emission lines.
}
\label{}
\end{figure}
We interpret this effect as indicative of 
galaxy-wide outflows which appear to be
a common characteristic of galaxies with large rates of star 
formation per unit area at high, as well as low, redshifts
(e.g. Heckman 2000).
Such `superwinds' involve comparable amounts of matter 
as is being turned into stars
(the mass outflow rate is of the same order as the star formation rate)
and about 10\% of the total kinetic energy delivered by the starburst
(Pettini et al. 2000).
Furthermore, they have a number of important astrophysical consequences.
They provide self-regulation to the star formation 
process (Efstathiou 2000);
can distribute the products of stellar
nucleosynthesis over large volumes
(the outflow speeds often exceed the escape 
velocities---Ferrara et al. 2000);
may account for some of the `missing' metals at high redshift
(Pettini 1999; Pagel 2000);
and may also allow Lyman continuum photons to leak from the 
galaxies into the intergalactic medium, easing the 
problem of how the universe came to be reionized.

\section{The Contribution of Galaxies to the Ionising Background}

Three of us (Steidel et al. 2001) have recently addressed
this last point quantitatively by adding together the spectra
of 29 Lyman break galaxies at a mean redshift
$\langle z \rangle = 3.40 \pm 0.09$ which places the Lyman
continuum region at wavelengths accessible to LRIS.
The composite spectrum, reproduced in Figure 6,
shows a positive signal at the $4.8 \sigma$ level
between rest-frame wavelengths 880 and 910\,\AA.
\begin{figure}
\vspace*{-1cm}
\hspace*{-0.8cm}
\centering
\includegraphics[width=0.8\textwidth,angle=270]{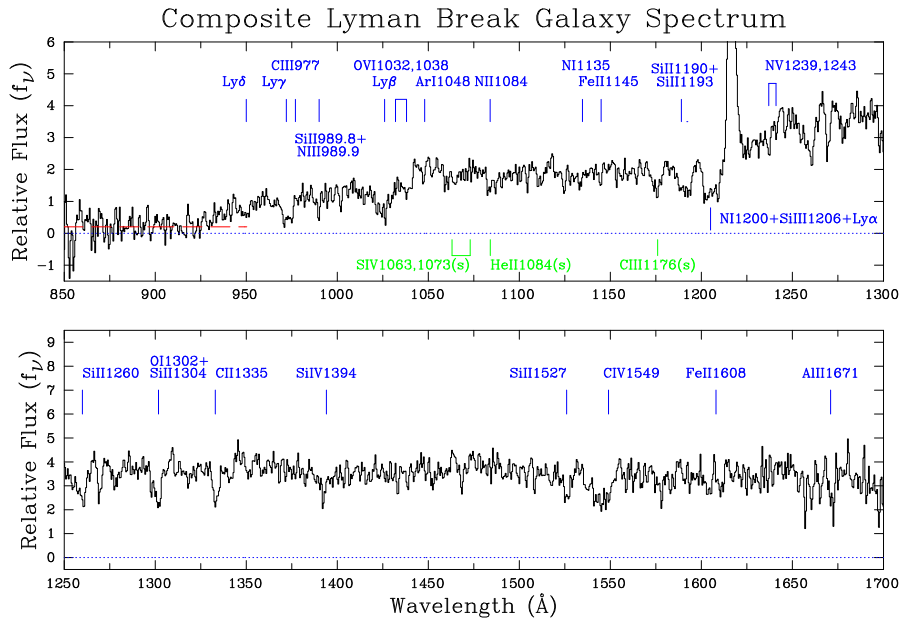}
\caption[]{Composite spectrum of 29 Lyman break galaxies at
$\langle z \rangle = 3.40$. Note the average residual flux 
in the Lyman continuum between 880 and 910\,\AA\ in the rest frame.
The most prominent interstellar
(blue and above the zero level) and stellar (green and below the zero 
level) absorption lines are identified.
}
\label{}
\end{figure}
After allowance for the opacity of the intergalactic medium,
even this weak signal implies a surprisingly large escape 
fraction---the emergent flux density below the Lyman limit is 20\%
of that at 1500\,\AA, which has been well quantified by
extensive surveys. This may well be an upper limit to
the value appropriate to the whole population because
the galaxies which contributed to the composite spectrum in Figure
6 are drawn from the bluest quartile of our full sample 
and may thus the most `leaky' objects among a spread of properties.
On the other hand, the ubiquitous presence of superwinds 
in Lyman break galaxies suggests that their interstellar media
may commonly develop large cavities which can provide a route
for the Lyman continuum photons to escape from the regions of star 
formation into the IGM.

If the escape fraction (as defined above) of 20\% applies
generally, the contribution of galaxies
to the metagalactic radiation field at $z \simeq 3.5$ is
$J_{\nu} \approx 1.2 \pm 0.3 
\times 10^{-21}$\,ergs~s$^{-1}$~cm$^{-2}$~Hz$^{-1}$~sr$^{-1}$.
This value is close to most estimates of $J_{\nu}$
based on the proximity effect in the \lya\ forest, and exceeds
the contribution from QSOs by about a factor of 5.
Presumably, the balance between galaxies and QSOs
as the providers of Lyman continuum photons shifts further
in favour of the former as we move to even higher redshifts
(Steidel et al. 1999; Fan et al. 2001), 
making it quite plausible that 
stars, rather than AGNs, were responsible for the end of the
`dark ages'. 
 
This conclusion, if supported by data soon to be secured
with the new UV-sensitive arm of LRIS, bodes well for the
{\it `Very Deep Fields'\/}¥ which are the goal of 
the Advanced Camera for Surveys 
on {\it HST\/}¥ and eventually the {\it NGST\/}¥.\\

Max Pettini would like to express his gratitude 
to the organisers of this enjoyable and timely meeting.

\clearpage
\addcontentsline{toc}{section}{Index}
\flushbottom
\printindex


\begin{thebibliography}{7}
%
\addcontentsline{toc}{section}{References}

\bibitem{} Bell, E.F. \& Kennicutt Jr., R.C. 2001, ApJ, in press
(astro-ph/0010340)

\bibitem{} Charlot, S. \& Longhetti, M. 2001, MNRAS, in press
(astro-ph/0101097)

\bibitem{} Dickinson, M. 2000, Philos. Trans. R. Soc. Lond. A, 358, 2001

\bibitem{} Efstathiou, G. 2000, MNRAS, 317, 697

\bibitem{} Fan, X., et al. 2001, AJ, in press (astro-ph/0008123)

\bibitem{} Ferguson, H.C., Dickinson, M., \& Williams, R. 2000, ARA\&A, 
38, 667

\bibitem{} Ferrara, A., Pettini, M., \& Shchekinov, Y. 2000,
MNRAS, 319, 539

\bibitem{} Flores, H., et al. 1999, ApJ, 517, 148

\bibitem{} Grevesse, N., \&  Sauval, A.J. 1998, Space Sci Rev, 85, 161

\bibitem{} Heckman, T.M. 2000, in ASP Conf. Ser.,
Gas and Galaxy Evolution,
ed. J.E. Hibbard, Mp.P. Rupen, \& J.H. van Gorkom,
(San Francisco:ASP), in press (astro-ph/0009075)

\bibitem{} Kennicutt Jr., R.C. 1998, ARA\&A, 36, 189

\bibitem{} Kobulnicky, H.A., \& Koo, D. 2001, ApJ, in press 
(astro-ph/0008242)


\bibitem{} Pagel, B.E.J. 2000, in Galaxies in the Young Universe,
ed. H. Hippelein (Berlin:Springer-Verlag), in press (astro-ph/9911204)

\bibitem{} Pettini, M. 1999, in Chemical Evolution from Zero to High 
Redshift, ed. J.R. Walsh, \& M.R. Rosa (Berlin:Springer-Verlag), 233

\bibitem{} Pettini, M. 2000, Philos. Trans. R. Soc. Lond. A, 358, 2035

\bibitem{} Pettini, M., Steidel, C.C., Adelberger, K.L., Dickinson, M.,
\& Giavalisco, M. 2000, ApJ, 528, 96


\bibitem{} Steidel, C.C., Adelberger, K.L., Giavalisco, M.,
Dickinson, M., \& Pettini, M. 1999, ApJ, 519, 1

\bibitem{} Steidel, C.C., Pettini, M., \& Adelberger, K.L. 2000, ApJ, 546
in press

\bibitem{} Sullivan, M., Treyer, M.A., Ellis, R.S., Bridges, T.J., 
Milliard, B., \& Donas, J. 2000, MNRAS, 312, 442

\bibitem{} Teplitz, H.I., et al. 2000, ApJ, 533, L65



\end{thebibliography}
\end{document}